
\font\twelverm=cmr10 scaled 1200    \font\twelvei=cmmi10 scaled 1200
\font\twelvesy=cmsy10 scaled 1200   \font\twelveex=cmex10 scaled 1200
\font\twelvebf=cmbx10 scaled 1200   \font\twelvesl=cmsl10 scaled 1200
\font\twelvett=cmtt10 scaled 1200   \font\twelveit=cmti10 scaled 1200

\skewchar\twelvei='177   \skewchar\twelvesy='60


\def\twelvepoint{\normalbaselineskip=12.4pt
  \abovedisplayskip 12.4pt plus 3pt minus 9pt
  \belowdisplayskip 12.4pt plus 3pt minus 9pt
  \abovedisplayshortskip 0pt plus 3pt
  \belowdisplayshortskip 7.2pt plus 3pt minus 4pt
  \smallskipamount=3.6pt plus1.2pt minus1.2pt
  \medskipamount=7.2pt plus2.4pt minus2.4pt
  \bigskipamount=14.4pt plus4.8pt minus4.8pt
  \def\rm{\fam0\twelverm}          \def\it{\fam\itfam\twelveit}%
  \def\sl{\fam\slfam\twelvesl}     \def\bf{\fam\bffam\twelvebf}%
  \def\mit{\fam 1}                 \def\cal{\fam 2}%
  \def\tt{\twelvett}
  \def\nullspace{\nulldelimiterspace=0pt \mathsurround=0pt }
  \def\big##1{{\hbox{$\left##1\vbox to 10.2pt{}\right.\nullspace$}}}
  \def\Big##1{{\hbox{$\left##1\vbox to 13.8pt{}\right.\nullspace$}}}
  \def\bigg##1{{\hbox{$\left##1\vbox to 17.4pt{}\right.\nullspace$}}}
  \def\Bigg##1{{\hbox{$\left##1\vbox to 21.0pt{}\right.\nullspace$}}}
  \textfont0=\twelverm   \scriptfont0=\tenrm   \scriptscriptfont0=\sevenrm
  \textfont1=\twelvei    \scriptfont1=\teni    \scriptscriptfont1=\seveni
  \textfont2=\twelvesy   \scriptfont2=\tensy   \scriptscriptfont2=\sevensy
  \textfont3=\twelveex   \scriptfont3=\twelveex  \scriptscriptfont3=\twelveex
  \textfont\itfam=\twelveit
  \textfont\slfam=\twelvesl
  \textfont\bffam=\twelvebf \scriptfont\bffam=\tenbf
  \scriptscriptfont\bffam=\sevenbf
  \normalbaselines\rm}



\def\beginlinemode{\endmode
  \begingroup\parskip=0pt \obeylines\def\\{\par}\def\endmode{\par\endgroup}}
\def\beginparmode{\endmode
  \begingroup \def\endmode{\par\endgroup}}
\let\endmode=\par
{\obeylines\gdef\
{}}
\def\singlespace{\baselineskip=\normalbaselineskip}
\def\oneandathirdspace{\baselineskip=\normalbaselineskip
  \multiply\baselineskip by 4 \divide\baselineskip by 3}
\def\oneandahalfspace{\baselineskip=\normalbaselineskip
  \multiply\baselineskip by 3 \divide\baselineskip by 2}
\def\doublespace{\baselineskip=\normalbaselineskip \multiply\baselineskip by 2}

\newcount\firstpageno
\firstpageno=2
\footline={\ifnum\pageno<\firstpageno{\hfil}\else{\hfil\twelverm\folio\hfil}\fi}
\let\rawfootnote=\footnote		
\def\footnote#1#2{{\rm\singlespace\parindent=0pt\rawfootnote{#1}{#2}}}
\def\raggedcenter{\leftskip=4em plus 12em \rightskip=\leftskip
  \parindent=0pt \parfillskip=0pt \spaceskip=.3333em \xspaceskip=.5em
  \pretolerance=9999 \tolerance=9999
  \hyphenpenalty=9999 \exhyphenpenalty=9999 }
\def\dateline{\rightline{\ifcase\month\or
  January\or February\or March\or April\or May\or June\or
  July\or August\or September\or October\or November\or December\fi
  \space\number\year}}
\def\received{\vskip 3pt plus 0.2fill
 \centerline{\sl (Received\space\ifcase\month\or
  January\or February\or March\or April\or May\or June\or
  July\or August\or September\or October\or November\or December\fi
  \qquad, \number\year)}}


\hsize=6.5truein
\vsize=8.9truein
\parskip=\medskipamount
\twelvepoint		
\doublespace		
\overfullrule=0pt	



\def\title			
  {\null\vskip 3pt plus 0.2fill
   \beginlinemode \doublespace \raggedcenter \bf}

\def\author			
  {\vskip 3pt plus 0.2fill \beginlinemode
   \singlespace \raggedcenter}

\def\affil			
  {\vskip 3pt plus 0.1fill \beginlinemode
   \oneandahalfspace \raggedcenter \sl}

\def\abstract			
  {\vskip 3pt plus 0.3fill \beginparmode
   \doublespace \narrower ABSTRACT: }

\def\endtitlepage		
  {\endpage			
   \body}

\def\body			
  {\beginparmode}		

\def\head#1{			
  \filbreak\vskip 0.5truein	
  {\immediate\write16{#1}
   \raggedcenter \uppercase{#1}\par}
   \nobreak\vskip 0.25truein\nobreak}

\def\refto#1{$^{#1}$}		

\def\references			
  {\head{References}		
   \beginparmode
   \frenchspacing \parindent=0pt \leftskip=1truecm
   \parskip=8pt plus 3pt \everypar{\hangindent=\parindent}}

\gdef\refis#1{\indent\hbox to 0pt{\hss#1.~}}	

\gdef\journal#1, #2, #3, 1#4#5#6{		
    {\sl #1~}{\bf #2}, #3, (1#4#5#6)}		

\gdef\journ2 #1, #2, #3, 1#4#5#6{		
    {\sl #1~}{\bf #2}: #3, (1#4#5#6)}		

\def\refstylenp{		
  \gdef\refto##1{ [##1]}				
  \gdef\refis##1{\indent\hbox to 0pt{\hss##1)~}}	
  \gdef\journal##1, ##2, ##3, ##4 {			
     {\sl ##1~}{\bf ##2~}(##3) ##4 }}

\def\refstyleprnp{		
  \gdef\refto##1{ [##1]}				
  \gdef\refis##1{\indent\hbox to 0pt{\hss##1)~}}	
  \gdef\journal##1, ##2, ##3, 1##4##5##6{		
    {\sl ##1~}{\bf ##2~}(1##4##5##6) ##3}}

\def\endreferences{\body}

\def\figurecaptions		
  {\endpage
   \beginparmode
   \head{Figure Captions}
}

\def\endpage			
  {\vfill\eject}

\def\endpaper			
  {\endmode\vfill\supereject}

\def\endit
  {\endpaper\end}


\def\ref#1{Ref. #1}			
\def\Ref#1{Ref. #1}			

\def\frac#1#2{{\textstyle #1 \over \textstyle #2}}

\def\sla{\raise.15ex\hbox{$/$}\kern-.57em}
\def\leaderfill{\leaders\hbox to 1em{\hss.\hss}\hfill}
\def\twiddle{\lower.9ex\rlap{$\kern-.1em\scriptstyle\sim$}}
\def\bigtwiddle{\lower1.ex\rlap{$\sim$}}
\def\gtwid{\mathrel{\raise.3ex\hbox{$>$\kern-.75em\lower1ex\hbox{$\sim$}}}}
\def\ltwid{\mathrel{\raise.3ex\hbox{$<$\kern-.75em\lower1ex\hbox{$\sim$}}}}
\def\square{\kern1pt\vbox{\hrule height 1.2pt\hbox{\vrule width 1.2pt\hskip 3pt
   \vbox{\vskip 6pt}\hskip 3pt\vrule width 0.6pt}\hrule height 0.6pt}\kern1pt}

\catcode`@=11
\newcount\r@fcount \r@fcount=0
\newcount\r@fcurr
\immediate\newwrite\reffile
\newif\ifr@ffile\r@ffilefalse
\def\w@rnwrite#1{\ifr@ffile\immediate\write\reffile{#1}\fi\message{#1}}

\def\writer@f#1>>{}
\def\referencefile{
  \r@ffiletrue\immediate\openout\reffile=\jobname.ref%
  \def\writer@f##1>>{\ifr@ffile\immediate\write\reffile%
    {\noexpand\refis{##1} = \csname r@fnum##1\endcsname = %
     \expandafter\expandafter\expandafter\strip@t\expandafter%
     \meaning\csname r@ftext\csname r@fnum##1\endcsname\endcsname}\fi}%
  \def\strip@t##1>>{}}

\def\citeall#1{\xdef#1##1{#1{\noexpand\cite{##1}}}}
\def\cite#1{\each@rg\citer@nge{#1}}	

\def\each@rg#1#2{{\let\thecsname=#1\expandafter\first@rg#2,\end,}}
\def\first@rg#1,{\thecsname{#1}\apply@rg}	
\def\apply@rg#1,{\ifx\end#1\let\next=\relax
\else,\thecsname{#1}\let\next=\apply@rg\fi\next}

\def\citer@nge#1{\citedor@nge#1-\end-}	
\def\citer@ngeat#1\end-{#1}
\def\citedor@nge#1-#2-{\ifx\end#2\r@featspace#1 
  \else\citel@@p{#1}{#2}\citer@ngeat\fi}	
\def\citel@@p#1#2{\ifnum#1>#2{\errmessage{Reference range #1-#2\space is bad.}%
    \errhelp{If you cite a series of references by the notation M-N, then M and
    N must be integers, and N must be greater than or equal to M.}}\else%
 {\count0=#1\count1=#2\advance\count1 by1\relax\expandafter\r@fcite\the\count0,%
  \loop\advance\count0 by1\relax
    \ifnum\count0<\count1,\expandafter\r@fcite\the\count0,%
  \repeat}\fi}

\def\r@featspace#1#2 {\r@fcite#1#2,}	
\def\r@fcite#1,{\ifuncit@d{#1}
    \newr@f{#1}%
    \expandafter\gdef\csname r@ftext\number\r@fcount\endcsname%
                     {\message{Reference #1 to be supplied.}%
                      \writer@f#1>>#1 to be supplied.\par}%
 \fi%
 \csname r@fnum#1\endcsname}
\def\ifuncit@d#1{\expandafter\ifx\csname r@fnum#1\endcsname\relax}%
\def\newr@f#1{\global\advance\r@fcount by1%
    \expandafter\xdef\csname r@fnum#1\endcsname{\number\r@fcount}}

\let\r@fis=\refis			
\def\refis#1#2#3\par{\ifuncit@d{#1}
   \newr@f{#1}%
   \w@rnwrite{Reference #1=\number\r@fcount\space is not cited up to now.}\fi%
  \expandafter\gdef\csname r@ftext\csname r@fnum#1\endcsname\endcsname%
  {\writer@f#1>>#2#3\par}}

\def\ignoreuncited{
   \def\refis##1##2##3\par{\ifuncit@d{##1}%
     \else\expandafter\gdef\csname r@ftext\csname r@fnum##1\endcsname\endcsname%
     {\writer@f##1>>##2##3\par}\fi}}

\def\r@ferr{\endreferences\errmessage{I was expecting to see
\noexpand\endreferences before now;  I have inserted it here.}}
\let\r@ferences=\references
\def\references{\r@ferences\def\endmode{\r@ferr\par\endgroup}}

\let\endr@ferences=\endreferences
\def\endreferences{\r@fcurr=0
  {\loop\ifnum\r@fcurr<\r@fcount
    \advance\r@fcurr by 1\relax\expandafter\r@fis\expandafter{\number\r@fcurr}%
    \csname r@ftext\number\r@fcurr\endcsname%
  \repeat}\gdef\r@ferr{}\endr@ferences}


\let\r@fend=\endpaper\gdef\endpaper{\ifr@ffile
\immediate\write16{Cross References written on []\jobname.REF.}\fi\r@fend}

\catcode`@=12

\citeall\refto		
\citeall\ref		%
\citeall\Ref		%
\catcode`@=11
\newcount\tagnumber\tagnumber=0

\immediate\newwrite\eqnfile
\newif\if@qnfile\@qnfilefalse
\def\write@qn#1{}
\def\writenew@qn#1{}
\def\w@rnwrite#1{\write@qn{#1}\message{#1}}
\def\@rrwrite#1{\write@qn{#1}\errmessage{#1}}

\def\taghead#1{\gdef\t@ghead{#1}\global\tagnumber=0}
\def\t@ghead{}

\expandafter\def\csname @qnnum-3\endcsname
  {{\t@ghead\advance\tagnumber by -3\relax\number\tagnumber}}
\expandafter\def\csname @qnnum-2\endcsname
  {{\t@ghead\advance\tagnumber by -2\relax\number\tagnumber}}
\expandafter\def\csname @qnnum-1\endcsname
  {{\t@ghead\advance\tagnumber by -1\relax\number\tagnumber}}
\expandafter\def\csname @qnnum0\endcsname
  {\t@ghead\number\tagnumber}
\expandafter\def\csname @qnnum+1\endcsname
  {{\t@ghead\advance\tagnumber by 1\relax\number\tagnumber}}
\expandafter\def\csname @qnnum+2\endcsname
  {{\t@ghead\advance\tagnumber by 2\relax\number\tagnumber}}
\expandafter\def\csname @qnnum+3\endcsname
  {{\t@ghead\advance\tagnumber by 3\relax\number\tagnumber}}

\def\equationfile{%
  \@qnfiletrue\immediate\openout\eqnfile=\jobname.eqn%
  \def\write@qn##1{\if@qnfile\immediate\write\eqnfile{##1}\fi}
  \def\writenew@qn##1{\if@qnfile\immediate\write\eqnfile
    {\noexpand\tag{##1} = (\t@ghead\number\tagnumber)}\fi}
}

\def\callall#1{\xdef#1##1{#1{\noexpand\call{##1}}}}
\def\call#1{\each@rg\callr@nge{#1}}

\def\each@rg#1#2{{\let\thecsname=#1\expandafter\first@rg#2,\end,}}
\def\first@rg#1,{\thecsname{#1}\apply@rg}
\def\apply@rg#1,{\ifx\end#1\let\next=\relax%
\else,\thecsname{#1}\let\next=\apply@rg\fi\next}

\def\callr@nge#1{\calldor@nge#1-\end-}
\def\callr@ngeat#1\end-{#1}
\def\calldor@nge#1-#2-{\ifx\end#2\@qneatspace#1 %
  \else\calll@@p{#1}{#2}\callr@ngeat\fi}
\def\calll@@p#1#2{\ifnum#1>#2{\@rrwrite{Equation range #1-#2\space is bad.}
\errhelp{If you call a series of equations by the notation M-N, then M and
N must be integers, and N must be greater than or equal to M.}}\else%
 {\count0=#1\count1=#2\advance\count1 by1\relax\expandafter\@qncall\the\count0,%
  \loop\advance\count0 by1\relax%
    \ifnum\count0<\count1,\expandafter\@qncall\the\count0,%
  \repeat}\fi}

\def\@qneatspace#1#2 {\@qncall#1#2,}
\def\@qncall#1,{\ifunc@lled{#1}{\def\next{#1}\ifx\next\empty\else
  \w@rnwrite{Equation number \noexpand\(>>#1<<) has not been defined yet.}
  >>#1<<\fi}\else\csname @qnnum#1\endcsname\fi}

\let\eqnono=\eqno
\def\eqno(#1){\tag#1}
\def\tag#1$${\eqnono(\displayt@g#1 )$$}

\def\aligntag#1$${\gdef\tag##1\\{&(\displayt@g##1 )\cr}\eqalignno{#1\\}$$
  \gdef\tag##1$${\eqnono(\displayt@g##1 )$$}}

\def\eqalignno#1{\displ@y \tabskip\centering
  \halign to\displaywidth{\hfil$\displaystyle{##}$\tabskip\z@skip
    &$\displaystyle{{}##}$\hfil\tabskip\centering
    &\llap{$\displayt@gpar##$}\tabskip\z@skip\crcr
    #1\crcr}}

\def\displayt@gpar(#1){(\displayt@g#1 )}

\def\displayt@g#1 {\rm\ifunc@lled{#1}\global\advance\tagnumber by1
        {\def\next{#1}\ifx\next\empty\else\expandafter
        \xdef\csname @qnnum#1\endcsname{\t@ghead\number\tagnumber}\fi}%
  \writenew@qn{#1}\t@ghead\number\tagnumber\else
        {\edef\next{\t@ghead\number\tagnumber}%
        \expandafter\ifx\csname @qnnum#1\endcsname\next\else
        \w@rnwrite{Equation \noexpand\tag{#1} is a duplicate number.}\fi}%
  \csname @qnnum#1\endcsname\fi}

\def\ifunc@lled#1{\expandafter\ifx\csname @qnnum#1\endcsname\relax}

\let\@qnend=\end\gdef\end{\if@qnfile
\immediate\write16{Equation numbers written on []\jobname.EQN.}\fi\@qnend}

\catcode`@=12


\input miniltx
\input graphicx.sty
\headline={\ifnum\pageno=1\firstheadline\else
\ifodd\pageno\rightheadline \else\leftheadline\fi\fi}
\def\firstheadline{\hfil}
\def\rightheadline{\hfil}
\def\leftheadline{\hfil}
        \footline={\ifnum\pageno=2\firstfootline\else\otherfootline\fi}
\def\firstfootline{\rm\hss\folio\hss}
\def\otherfootline{\hfil}
\firstpageno=2
\footline={\ifnum\pageno<\firstpageno{\hfil}\else{\hfil\twelverm\folio\hfil}\fi}

\font\twelvebf=cmbx10 scaled\magstep 1
\font\twelverm=cmr10 scaled\magstep 1
\font\twelveit=cmti10 scaled\magstep 1

\font\tenbf=cmbx10
\font\tenrm=cmr10

\parindent=1.2pc
\hsize=6.0truein
\vsize=8.6truein
\def\(#1){(\call{#1})}
\def\call#1{{#1}}


\def\references{{\noindent\bf 9.
References}\beginparmode\frenchspacing
               \leftskip=1truecm\parskip=8pt plus 3pt
                \everypar{\hangindent=3em}}

\line{}
\vskip .26 in

\centerline{\bf QUANTUM MECHANICS AT THE PLANCK SCALE\footnote{$^*$}{Talk
given at the {\sl Workshop on Physics at the Planck Scale}, Puri, India,
December 12--21, 1994. This talk is a pr\'ecis of the author's 1992 Les
Houches Lectures {\it Spacetime Quantum Mechanics and the Quantum
Mechanics of Spacetime}, Ref [\cite{Harpp}].}}
\vskip .50 in
\centerline{JAMES B.~HARTLE\footnote{$^{**}$}{e-mail:
hartle@cosmic.physics.ucsb.edu}}
\baselineskip=12pt
\centerline{\it Department of Physics, University of California}
\baselineskip=10pt
\centerline{\it Santa Barbara, CA 93106-9530}
\vskip .13 in
{
\midinsert\narrower\narrower
\font\tenrm=cmr10
\vskip .13 in
\centerline{ABSTRACT}
\noindent
{\tenrm Usual quantum mechanics requires a fixed, background, spacetime
geometry and its associated causal structure. A generalization of the
usual theory may therefore be needed at the Planck scale for quantum
theories of gravity in which spacetime geometry is a quantum variable.
The elements of generalized quantum theory are briefly reviewed and
illustrated by generalizations of usual quantum theory that
incorporate spacetime alternatives, gauge degrees of freedom, and
histories that move forward and backward in time. A generalized quantum
framework for cosmological spacetime geometry is sketched. This theory
is in fully four-dimensional form and free from the need for a fixed
causal structure. Usual quantum mechanics is recovered as an
approximation to this more general framework that is appropriate in
those situations where spacetime geometry behaves classically.}
\endinsert
}

\oneandathirdspace
\vskip .13 in
\taghead{1.}
\leftline{\bf 1. Introduction}

Generalizations of usual quantum theory 
are required for physics at the Planck Scale. That is because the usual
framework depends strongly on an assumed fixed, background, spacetime
geometry. States are defined on spacelike surfaces in this geometry.
States evolve unitarily in between such surfaces in the absence of
measurement and by state vector reduction on them when a measurement
occurs. The inner product between states is defined by integrals over
fields on a spacelike surface. These are just some of the ways that a 
fixed  spacetime
geometry is central to the usual formulation of quantum mechanics.
However, at the Planck scale, spacetime geometry is not fixed, but a
quantum dynamical variable, --- fluctuating and without definite value.
Given two nearby points on a spacetime manifold it is not possible to
say whether they are spacelike separated or not. Rather, the amplitudes
for prediction are sums over {\it different} metrics on the manifold.
Points separated by a spacelike interval in
one metric may be timelike separated in another that contributes just as
significantly to the sum. For this reason, usual quantum mechanics needs to be
generalized to accommodate quantum spacetime. There are many approaches
to this generalization whose difficulties have been lucidly reviewed 
in\refto{Kuc92,Ish92,Ish93,Unr91}. This lecture describes another 
approach --- the effort to provide a generalization of usual 
quantum mechanics that is in fully spacetime form and
does not require a fixed spacetime geometry but
which yields the usual formulation approximately in those physical
situations where spacetime geometry {\it is} approximately fixed.

The rules for the calculation of $S$-matrix elements in field theory may
be summarized without reference to states on spacelike surfaces except
in asymptotic regions. It is thus possible to carry out investigations
of the dynamics of quantum gravity as expressed by $S$-matrix elements
without addressing the issues of generalization raised above. However,
one of the principal applications of quantum gravity is to the very
early universe where the physics of the Planck scale becomes important.
In quantum cosmology we cannot evade the challenge of generalizing
quantum theory. We live in the middle of this particular experiment.

The application of quantum mechanics to cosmology also  requires another
kind of generalization of the usual formulation.
Usual quantum mechanics predicts 
the outcomes of ``measurements'' carried out on a system by another
system outside it. But in cosmology there is no system outside.
Cosmology requires a quantum mechanics of closed systems that is a
generalization of the usual theory. Recent years have seen the emergence of
such a generalization built on the work of Everett, Zeh, Zurek,
Griffiths, Omn\`es, Gell-Mann, and many 
others.\footnote{$^{\rm a}$}{For an introduction from the author's 
perspective see Ref.~[\cite{Harsum}].}
  The most general
predictions of this formulation of quantum mechanics are the
probabilities of the individual members of
 sets of alternative histories of the closed
system. Consistency of probability sum rules is the criterion
determining the sets of histories which may be assigned probabilities rather
than any notion of measurement. The absence of quantum mechanical
interference between histories, or decoherence, is the sufficient
condition for this consistency. The initial condition of the closed
system and Hamiltonian determine which sets of histories decohere rather
than the action of any external observer. 

There is a relation between these two required generalizations.
By abstracting the principles of the quantum
mechanics of closed systems one arrives at a more general framework for
quantum prediction --- generalized quantum mechanics. Within generalized
quantum mechanics one can construct fully spacetime quantum theories
which do not require a fixed background spacetime. In this lecture we
shall review some simple examples of generalized quantum theories and
sketch a generalized quantum mechanics of spacetime geometry. Our
discussion is necessarily brief; for a more extensive one see
Ref.~[\cite{Harpp}].\footnote{$^{\rm b}$}{This lecture may be thought of as an
abridgement of  the author's lectures in Ref.~[\cite{Harpp}].} 
\vskip .13 in
\taghead{2.}
\leftline{\bf II. Generalized Quantum Theory}

Not every
set of histories that may be described can be consistently assigned
probabilities in quantum theory. In the two slit experiment
 it would be inconsistent to assign probabilities to the
alternatives that the electron went through slit $A$ or slit $B$ on its way
to detection at a point $y$ on the screen in the 
absence of a
measurement of which happened. The probability to arrive at $y$
would not be the sum of the probability to go through $A$ and arrive at $y$,
and the probability to go through $B$ and arrive at $y$, because of quantum
interference. In quantum mechanics probabilities are squares of
amplitudes, and
$$
\bigl|\psi_A(y) + \psi_B(y)\bigr|^2  \not=\bigl|\psi_A(y)\bigr|^2 +
\bigl|\psi_B(y)\bigr|^2\ . \tag twoone
$$
A criterion is therefore needed to specify which sets of histories may
be consistently assigned probabilities. In the familiar quantum
mechanics that criterion is {\it measurement} --- probabilities may be
consistently assigned to histories of {\it measured} alternatives and
not in general otherwise.  In the two slit experiment if we measure
which slit the electron passes through, then interference is destroyed, the
probability sum rules are obeyed, and probabilities are predicted for
the two alternative histories of the electron.

However, a criterion based on measurements or observers cannot be
fundamental in a quantum theory that seeks to explain the early universe
where neither existed. A more general criterion for closed systems
assigns probabilities to just those sets of histories for which there is
vanishing interference between its individual members as a consequence of
the system's initial quantum state\refto{Gri84,Omnsum,GH90a}. Such sets
of histories are said to {\it decohere}. Decoherent sets of histories
are what may be used for prediction and retrodiction in quantum
cosmology for they may be assigned probabilities.

The usual Hamiltonian formulation of quantum mechanics in a fixed
background spacetime was the context for the treatment of the decoherent
sets of histories of a closed system in Refs [\cite{Gri84,Omnsum,GH90a}]. 
However, it is possible to abstract from these discussions principles
for a wider class of quantum mechanical theories called {\it generalized
quantum theories}. Within that framework, generalized quantum theories of
spacetime geometry may be constructed which do not assume a fixed
background spacetime.

The principles of generalized quantum mechanics were introduced in
Ref.~[\cite{Har91a}] and developed more fully in Refs [\cite{Har91b}] and
[\cite{Harpp}].
The principles have been axiomatized in a rigorous
mathematical setting by Isham\refto{Ish94}. Three elements are needed
to specify a generalized quantum theory:

\item{(1)} The sets of {\it fine-grained histories}. These are the most
refined possible description of a closed system. 
\item{(2)} The allowed {\it coarse grainings}. A coarse graining of a
set of histories is
generally a partition of that set into 
mutually exclusive classes $\{c_\alpha\}, \alpha=1,2, \cdots$ called
{\it coarse-grained histories}. Each coarse-grained history is 
a set of fine-grained histories and the set of classes constitutes
a set of coarse-grained histories with each history labeled by the
discrete index $\alpha$. 
\item{(3)} A {\it decoherence functional} defined for each allowed set of
coarse-grained histories which incorporates a theory of the initial
condition and dynamics of the closed system and measures the quantum 
mechanical
interference between pairs of histories in the set. A decoherence
functional $D(\alpha^\prime,\alpha)$ must satisfy the following
properties.
\itemitem{(i)} {\sl Hermiticity}: 
$$
D(\alpha^\prime,\alpha)= D^*(\alpha,
\alpha^\prime)\ .\tag twotwo a
$$
\itemitem{(ii)} {\sl Positivity}: 
$$
D(\alpha, \alpha) \geq 0\ . \tag twotwo b
$$
\itemitem{(iii)} {\sl Normalization}:
$$
\sum\nolimits_{\alpha^\prime\alpha} D(\alpha^\prime, \alpha) = 1\ .
\tag twotwo c
$$
\itemitem{(iv)} {\sl The Principle of Superposition}:\hfill\break
If $\{\bar c_{\bar\alpha}\}$
 is a coarse graining of a set of histories
$\{c_\alpha\}$, that is, a further partition into classes $\{\bar
c_{\bar\alpha}\}$, then
$$
D(\bar\alpha^\prime, \bar\alpha) =
\sum\limits_{\alpha^\prime\in\bar\alpha^\prime}
\ \sum\limits_{\alpha\in\bar\alpha} D(\alpha^\prime, \alpha)\ .
\tag twotwo d
$$

Once these three elements are specified the process of prediction
proceeds as follows: A set of histories is said to (medium) decohere if
all the ``off-diagonal'' elements of $D(\alpha^\prime, \alpha)$ are
sufficiently small. The diagonal elements are the probabilities
$p(\alpha)$ of the individual histories in a decoherent set. These two
definitions are summarized in the one relation
$$
D(\alpha^\prime, \alpha) \approx \delta_{\alpha^\prime\alpha}
p(\alpha)\ . \tag twothree
$$
As a consequence of \(twothree) and the four properties of \(twotwo), the
numbers $p(\alpha)$ lie between zero and one, sum to one, and satisfy
the most general form of the probability sum rules
$$
p(\bar\alpha) = \sum\limits_{\alpha\in\bar\alpha} p(\alpha)
\tag twofour
$$
for any coarse graining $\{\bar c_{\bar\alpha}\}$ of the set
$\{c_\alpha\}$. The $p(\alpha)$ are therefore probabilities. They are
the predictions of generalized quantum mechanics for the possible
coarse-grained histories of the closed system that arise from the 
theory of its initial
condition and dynamics incorporated in the construction of $D$.

In the following we shall illustrate this general framework with
examples designed to realize it in concrete form but also designed
to show how to cast quantum theories into fully spacetime form. For
these illustrative purposes we shall confine ourselves to
sum-over-histories formulations which posit a unique set of fine-grained
histories. Relations to operator formulations are discussed in Ref.~[\cite
{Harpp}].

\vskip .13 in
\taghead{3.}
\leftline{\bf 3. Non-Relativistic Quantum Mechanics}

The non-relativistic quantum mechanics of a particle moving in one
dimension is the simplest example of generalized quantum theory when 
its three elements are identified as follows:

\item{(1)} {\sl Fine-Grained Histories}.
The fine-grained histories are particle paths $x(t)$ which are {\it
single-valued} functions of $t$ on a fixed time interval, say, $[0, T]$.

{\item{(2)} {\sl Allowed Coarse Grainings}. The fine-grained paths may be
partitioned by their behavior with respect to regions of $x$ at definite
moments of time. For instance, consider a division of the real line into
intervals $\{\Delta^1_{\alpha_1}\}$ at time $t_1$ and
$\{\Delta^2_{\alpha_2}\}$ at time $t_2$ where $\alpha_1$ and $\alpha_2$
run over a discrete set of labels. One coarse-grained history consists
of all the paths which pass through, say, $\Delta^1_8$and $\Delta^2_{17}$.
This is the coarse-grained history in which the particle is localized in
region $\Delta^1_8$ at time $t_1$ and $\Delta^2_{17}$ at time $t_2$. An
exhaustive set of coarse-grained histories in this example
is specified by considering
all possible sequences of two regions in the sets $\{\Delta^1_{\alpha_1}\}$
and $\{\Delta^2_{\alpha_2}\}$.
\item{3.} {\sl Decoherence Functional}. Given a partition of the set of
fine-grained paths into classes $\{c_\alpha\}, \alpha = 1,2,\cdots$ we
may define a class operator for each class (coarse-grained history) by
a sum over paths in the class, {\it viz.}:
$$
\bigl\langle x^{\prime\prime} \bigl | C_\alpha \bigr |x^\prime \bigr\rangle
= \int_{c_\alpha} \delta x\, \exp \bigl(iS[x(\tau)]\bigr)\ .
\tag threeone
$$
The functional $S[x(\tau)]$ is the classical action which defines
quantum dynamics in path integrals. The sum is over
paths which start at $x^\prime$ at $t=0$, proceed to $x^{\prime\prime}$
at $t=T$ and lie in the class $c_\alpha$. A sum of the form \(threeone)
over the class of {\it all} paths is just an expression for the propagator 
$\langle x^{\prime\prime}|\exp (-iHT)|x^\prime\rangle$. Thus,
$$
\sum\limits_\alpha C_\alpha = e^{-iHT}\ .
\tag threetwo
$$
With the class operators in hand, the decoherence functional for
non-relativistic quantum mechanics is
$$
D(\alpha^\prime, \alpha) = Tr\bigl[C_{\alpha^\prime} \rho
\, C^\dagger_\alpha\bigr]
\tag threethree
$$
where $\rho$ is the density matrix representing the initial condition of
the closed system. It is not difficult to see from \(threeone),
\(threetwo), and the properties of density matrices, that the four
requirements decoherence functionals \(twotwo) are  satisfied.

The class operators are simply expressed for the coarse grainings by
exhaustive sets of ranges of positions at definite movements of time
that we have taken to {\it define} non-relativistic quantum mechanics
above. If $\{P^k_{\alpha_k} (t_k)\}$ are a set of Heisenberg-picture
projections onto ranges $\{\Delta^k_{\alpha_k}\}$ at time $t_k$, then
the class operator for the paths that pass through intervals
$\Delta^1_{\alpha_1}, \cdots, \Delta^n_{\alpha_n}$ at $t_1\cdots, t_n$
is
$$
C_{\alpha_n\cdots\alpha_1} = e^{-iHT} P^n_{\alpha_n}(t_n) \cdots
P^1_{\alpha_1}(t_1)\ .
\tag threefour
$$
In \(threefour) one can see the two forms of evolution in usual quantum
mechanics. Applied to a state, \(threefour) evolves it unitarily in
between times $t_1,\cdots, t_n$ (Heisenberg-picture evolution of the
$P$'s) and by state vector reduction (action of the $P$'s) at the times
$t_1,\cdots, t_n$.

This formulation of non-relativistic quantum mechanics is not in fully
spacetime form. Quantum dynamics has been expressed in spacetime form
through the use of Feynman's path integrals over histories. But the
alternatives are restricted to sequences of sets of alternative ranges
of position at {\it
definite moments of time}. However, this usual framework of
non-relativistic theory is easily generalized to alternatives that are
in spacetime form. Simply allow arbitrary
partitions as coarse grainings and replace (2) by 

\item{($2^\prime$)} {\sl Allowed Coarse Grainings}: Arbitrary partitions of the
fine-grained paths.

This allows new  types of {\it spacetime} alternatives which are
not at definite moments of time.  For example, given a region $R$ with
extent in both space and time, one could partition the paths into the
class $c_0$ of paths that never cross $R$ and the class $c_1$ that cross
$R$ sometimes.
A way to see that this is a genuine generalization is to note
that the resulting class operators are generally not unitary and neither
are they projections or products of projections as in \(threefour). 
Thus, this generalized
quantum mechanics of a non-relativistic particle cannot be reformulated in
terms of states on surfaces of constant time which evolve
unitarily or by reduction. Even without states on spacelike surfaces the
theory is predictive as described above and this generalization is now
in fully spacetime form with respect to both dynamics and alternatives.

\vskip .13 in
\taghead{4.}
\leftline{\bf 4. Abelian Gauge Fields}

The three elements of a generalized quantum mechanics of the free
electromagnetic field are:

\item{(1)} {\sl Fine-Grained Histories}: Configurations of the potential
$A_\mu(x)$ on a spacetime region between two constant time surfaces,
say, those at $t=0$ and $t=T$.

\item{(2)} {\sl Allowed Coarse-Grainings}: Partitions of the fine-grained
histories of the potential into {\it gauge-invariant} classes
$\{c_\alpha\}$.

\item{(3)} {\sl Decoherence Functional}: Class operators on the physical
Hilbert space of transverse degrees of freedom of the vector potential
are defined by
$$
\bigl\langle\vec A^{T^{\prime\prime}}\bigl| C_\alpha \bigr| \vec
A^{\prime T}\bigr\rangle = \int_{c_\alpha} \delta A\, \Delta_\Phi [A]
\,\delta [\Phi(A)]\, \exp \bigl(iS[A\bigr])\ .
\tag fourone
$$
The functional $S[A]$ is the action for the free 
electromagnetic field.
$\Phi(A)$ is a function such that $\Phi(A)=0$ is a gauge fixing
condition, and $\Delta_\Phi[A]$ is the associated Faddeev-Popov
determinant. The sum is over all potentials $A_\mu(x)$ which match $\vec
A^{T^\prime}({\bf x})$ on the initial surface $t=0$ and $\vec
A^{T^{\prime\prime}}({\bf x})$ on the final surface $t=T$. In particular
the values of the time component
 $A^t({\bf x})$ and the longitudinal component of $\vec
A({\bf x})$ on these surfaces are summed over. The decoherence
functional is 
$$
D(\alpha^\prime, \alpha) = Tr\bigl[C_{\alpha^\prime} \rho\,
C^\dagger_\alpha\bigr]
\tag fourtwo
$$
where all operators and operations are defined in the Hilbert space of
the transverse components of the vector potential.

When restricted to partitions by values of $\vec A^T(x)$ on
spacelike surfaces, the generalized quantum mechanics specified above
coincides with the familiar Hamiltonian quantum mechanics of the
electromagnetic field. But more general alternatives are possible. One
could consider partitions by the values of any gauge invariant
functional, for instance the average of a magnetic field component over a
spacetime region $R$
$$
F[A] = \frac{1}{\Delta V} \int_R d^4 x  B_z(x)\ .
\tag fourthree
$$
More specifically let $\{\Delta_\alpha\}$ be a set of mutually exclusive
intervals making up the real line. The coarse-grained history $c_\alpha$
consists of all those potentials $A_\mu(x)$ for which $F[A]\in
\Delta_\alpha$. Such field averages are familiar from Bohr and
Rosenfeld's discussion of the measurability of the electromagnetic
field.

Beyond functionals of the type \(fourthree) which depend only on the
transverse degrees of freedom one can also consider partitions by values
of averages of quantities like $\vec\nabla\cdot \vec E$ which are gauge
invariant but involve non-transverse degrees of freedom. In this way the
question of whether the constraint $\vec \nabla\cdot \vec E =0$ is
satisfied becomes a question of the probability of its value rather than
a matter of the absence of the relevant degrees of freedom.

\vskip .13 in
\taghead{5.}
\leftline{\bf 5. The Relativistic World Line}

Classical dynamics in general relativity may be thought of as the
evolution of the spatial geometry of a family of spacelike surfaces that
foliate spacetime. This dynamics may be exhibited by decomposing the
metric in $3+1$ form with respect to these surfaces:
$$
ds^2 = - N^2 dt^2 + h_{ij} \bigl(dx^i + N^i dt\bigr) \bigl(dx^i + N^i
dt\bigr)\ .
\tag fiveone
$$
Here, $t$ is the label of a spacelike surface and $x^i$ are three
co\"ordinates in it.

Two types of invariance of the action of general relativity may be
distinguished when it is expressed in terms of the variables defined by
\(fiveone). First there are transformations of the co\"ordinates $x^i$
in the spacelike surfaces
$$
x^i \to \bar x^i(x^j)\ , \tag fivetwo
$$
which for infinitesimal transformations $x^i\to x^i + \xi^i(x^j)$ leads to
$$
h_{ij}(x)\to h_{ij}(x) + D_{(i}\xi_{j)}(x)\ . \tag fivethree
$$
where $D_i$ is the derivative in the surface.
The transformation rule \(fivethree) is not unlike that of gauge
transformations in electromagnetism. It can be treated in a way
analogous to that of Section 4 in the construction of a generalized
quantum framework for general relativity.

A class of invariances of a different character are the
reparameterizations of the time
$$
t\to \bar t(t)\ , \tag fivefour
$$
and these have their own special implications for the construction of a
generalized quantum mechanics for gravitation. A simple model theory
which exhibits time reparameterization invariance classically is the
relativistic world line. Its action may be taken to be
$$
S\left[x^\mu, N\right] = \frac{m}{2}\int\nolimits^1_0 d\lambda\, N(\lambda)
\left[\left(\frac{\dot x^\mu(\lambda)}{N(\lambda)}\right)^2-1\right]\ .
\tag fivefive
$$
Here, $x^\mu(\lambda)$ are the co\"ordinates of the world line as
functions of a parameter $\lambda$ along it, a dot denotes a derivative
with respect to $\lambda$, and $N(\lambda)$ is a multiplier enforcing
the constraint $(p^\mu)^2 = -m^2$. This action is invariant under
reparameterizations 
$\lambda\to\bar\lambda(\lambda)$ that leave the endpoints fixed, with an
appropriate transformation law for $N$.

It is not possible to construct a quantum theory of a {\it single}
relativistic particle interacting with an external field within the
framework of usual Hamiltonian quantum mechanics. That is because the
external field will produce pairs so that the only consistent theory is
a {\it many} particle theory. However, it {\it is} possible to construct
a generalized quantum theory of a single relativistic world line even
interacting with an external field. This is not a theory of realistic
relativistic particles such as the proton and electron for that is
provided by field theory. However, in many ways this is a useful
model for the quantum cosmology of a universe with a single fixed
topology. We now sketch the elements of a generalized quantum theory of
a single relativistic world line confining ourselves for simplicity to
the non-interacting case.

\item{(1)} {\sl Fine-grained histories}: The fine-grained histories are
paths $(x^\mu(\lambda), N(\lambda))$ in the extended configuration space
of paths in spacetime $x^\mu(\lambda)$ and multiplier $N(\lambda)$. The
paths in spacetime are not single valued in the time of any Lorentz
frame, but move both forward and backward in the time of any Lorentz
frame.
\item{(2)} {\sl Allowed Coarse Grainings}: Any partition of the
fine-grained histories into {\it repara-} {\it metrization invariant} classes
$c_\alpha$ is an allowed coarse graining. For example, given a fixed
spacetime region $R$, one could partition the paths $x^\mu(\lambda)$
into those which never cross $R$ and the class which cross $R$ at least
once. One could partition the fine-grained histories between two such
regions by the value of the reparameterization invariant ``proper time''
$$
\int N(\lambda) d\lambda
\tag fivesix
$$
between the last passage of the first region and the first passage of
the second. More generally, one could partition the fine-grained
histories by the values of any reparameterization invariant functional
$F[x^\mu(\lambda), N(\lambda)]$.

\item{}It is {\it not} possible to partition this class of paths by the 
position that the world line intersects
 a spacelike surface. The world line may
cross a given surface, not just at one position, but at an arbitrarily
large number of them.

\item{(3)} {\it Decoherence Functional}: For each class of fine-grained
histories $c_\alpha$ in an allowed coarse-graining, the matrix elements
of a class operator may be defined by
$$
\left\langle x^{\prime\prime}\left\Vert C_\alpha\right\Vert
x^\prime\right\rangle = \int\nolimits_{c_\alpha} \delta x\, \delta N\,
\Delta_\Phi [x,N]\, \delta\, [\Phi(x,N)]\ \exp(iS[x,N])\ .
\tag fiveseven
$$
Here, the sum is over all fine-grained histories whose paths start at
the spacetime point $x^{\prime\prime\mu}$, end at the point
$x^{\prime\mu}$, and lie in the class $c_\alpha$ The functional $S$ is
the action \(fivefive). The condition $\Phi=0$ fixes a parameterization
and $\Delta_\Phi$ is the associated Faddeev-Popov determinant. The
measure for the integration is that induced by the Liouville ``$dpdq$''
measure on phase space.

Initial and final conditions are represented by wave functions $\psi(x)$
and $\phi(x)$ on initial and final spacelike surfaces in spacetime,
labeled $\sigma^\prime$ and $\sigma^{\prime\prime}$ respectively. We
define class operators in linear spaces of such wave functions
$\{\psi_j(x)\}$ and $\{\phi_i(x)\}$ by ``attaching'' wave functions
to \(fiveseven) using a bilinear (but not necessarily positive) inner
product $\circ$, {\it viz.}:
$$
\left\langle\phi_i\left|C_\alpha\right|\psi_j\right\rangle =
\phi^*_i(x^{\prime\prime})\, \circ
\left\langle x^{\prime\prime}\left\Vert C_\alpha\right\Vert
x^\prime\right\rangle\, \circ\, \psi_j (x^\prime)\ .
\tag fiveeight
$$
The Klein-Gordon inner product is an appropriate choice for $\circ$
 for the relativistic world
line.  Specifically, in the case that $\sigma^\prime$ and
$\sigma^{\prime\prime}$ are surfaces of constant time in some Lorentz
frame, we define
$$
\left\langle\phi_i\left|C_\alpha\right|\psi_j\right\rangle =
-\int_{\sigma^{\prime\prime}} d^3 x^{\prime\prime} \int_{\sigma^\prime}
d^3x^\prime \phi^*_i(x^{\prime\prime})
\frac{\buildrel\leftrightarrow\over\partial}{\partial t^{\prime\prime}} 
\left\langle x^{\prime\prime}
\left\Vert C_\alpha\right\Vert x^\prime\right\rangle
\frac{\buildrel\leftrightarrow\over\partial}{\partial t^\prime} 
\psi_j(x^\prime)\ .
\tag fivenine
$$
If the wave functions defined off the surfaces $\sigma^\prime$ and
$\sigma^{\prime\prime}$ are consistent with the operator form of the
constraint $(p^\mu)^2 = - m^2$, i.e.~satisfy the Klein-Gordon equation:
$$
\bigl(-\nabla^2 + m^2\bigr) \phi_i (x) = 0\ ,
\tag fiveten
$$
then the construction \(fivenine) is independent of deformations of
$\sigma^\prime$ and $\sigma^{\prime\prime}$ provided they do not
intersect.

For an initial condition represented by a single wave function $\psi(x)$,
and a final condition of equal probability for a set of wave functions
$\{\phi_i(x)\}$,  the decoherence functional is
$$
D(\alpha^\prime, \alpha) = {\cal N} \sum_i
\left\langle\phi_i\left|C_{\alpha^\prime}\right| \psi \right\rangle
\left\langle\psi\left|C_\alpha\right|\phi_i\right\rangle
\tag fiveeleven a
$$
where ${\cal N}$ is a normalizing factor
$$
{\cal N}^{-1} = \sum_i
\left|\left\langle\phi_i\left|C_u\right|\psi\right\rangle\right|^2
\ ,
\tag fiveeleven b
$$
with $C_u$ being the class operator for {\it all} fine-grained
histories. It is not difficult to verify that this decoherence
functional satisfies the four general requirements \(twotwo).

With the decoherence functional \(fiveeleven), sets of alternative histories
which decohere may be identified and the probabilities of the individual
alternative histories computed. This generalized quantum mechanical
framework cannot be reformulated in terms of states on spacelike
surfaces, their unitary evolution and reduction. The inclusion of paths
that move both forward and backward in time implies it is not meaningful
to consider alternative values of position at moments of time so that
identities like \(threefour) can no longer hold. Thus, by moving beyond the
strictures of usual Hamiltonian quantum mechanics, to a more general
quantum framework, we have exhibited a quantum theory of a single
relativistic world line in fully spacetime form.

\vskip .13 in
\taghead{6.}
\leftline{\bf 6. General Relativity}

The question of whether there is a consistent manageable quantum theory of
Einstein's general relativity is still open. Whatever the answer, a
generalized quantum theory of general relativity provides a 
model of the kind of conceptual issues that must be faced in any quantum
theory of gravity.  Further, such a generalized quantum theory, made
finite by truncating its ultraviolet divergences, may be a useful tool in
quantum cosmology for investigating the predictions of a theory of the
initial condition for the very low energy phenomena of the universe like the
galaxy-galaxy correlation function.

As mentioned above, the classical theory of general relativity exhibits
symmetries analogous both to gauge symmetries and reparameterizations of
time. Building on the treatments of gauge theories in Section 4 and the
relativistic world line in Section 5 it is possible to sketch the elements of a
generalized quantum theory for quantum spacetime. For simplicity we
restrict attention to spatially closed cosmological four-geometries.

\item{(1)} {\sl Fine-grained histories}. A class of metrics on four
dimensional manifolds are the fine-grained histories of a generalized
quantum theory of spacetime geometry. The framework is broad enough to
allow for different manifolds and thus discuss topology change, but for
simplicity in this discussion we restrict attention to the case where
the manifold is fixed and of form ${\bf R}\times M^3$ where $M^3$ is a
closed three manifold
that supports spatially closed cosmological geometries. What
behavior is permitted the fine-grained histories
 on very small scales, and what singularities are
allowed on large scales, are two issues related to the ultraviolet
behavior of the theory that we shall not discuss. The geometries should
at least be such that the action for general relativity is finite.

\item{2.} {\sl Allowed Coarse Grainings}. The general idea is a
partition of the fine-grained metrics into four-dimensional
{\it diffeomorphism-invariant}
classes $\{c_\alpha\}$. For example, we could partition all cosmological
four metrics into the class that contain no three surface with a volume
bigger than $V_0$ and the class that contain at least one such three
surface. The probability of the second of these alternatives can be
thought of as the probability that the universe expands to a volume
bigger than $V_0$.
The set of fine-grained histories could be partitioned 
into the class of
geometries which is homogeneous and isotropic to some standard at large
volumes and the class of
 geometries which is not homogeneous and isotropic to that
standard. The probability of the first class is the probability that the
universe becomes homogeneous and isotropic far from its singularities.
The set of fine-grained histories could be partitioned into geometries
which obey the Einstein equations to some accuracy at large volumes and
the class which do not obey the Einstein equation. The probability of
the first class is the probability that the universe becomes classical
at large size. With further partitions of this class the probabilities
of individual classical histories could be calculated.

\item{} In general when we ask  for the probability of {\it any}
property of the universe which can be expressed in terms of spacetime
geometry and matter fields there is a corresponding partition of the
fine-grained histories into the class which has this property and the
class which does not have it. If it is not possible to tell which
four-dimensional geometries have the property, and which do  not, then the
property is not well defined. Coarse grainings by partitions of the
fine-grained histories into diffeomorphism-invariant classes is thus the
most general notion in quantum theory
 of a set of alternatives expressible in terms of
spacetime geometry.

\item{(3)} {\it Decoherence Functional}. The construction of a
decoherence functional for general relativity parallels the construction
of that for the relativistic particle. We sketch it here referring the
reader to [\cite{Harpp}] for more details. Consider histories on a
manifold $M={\bf I} \times M^3$ with two end boundaries $\partial M^\prime$
and $\partial M^{\prime\prime}$. Any four-dimensional metric and matter
field configuration on $M$ induces a three-metric $h^\prime_{ij} ({\bf x})$
and a field configuration $\chi^\prime ({\bf x})$ on $\partial M^\prime$. 
Similarly 
three-metrics and fields are induced on the other end of the history
$\partial M^{\prime\prime}$. Class operator matrix elements for a 
class $c_\alpha$ may be defined by
$$
\bigl\langle h^{\prime\prime}_{ij}, \chi^{\prime\prime}\left\Vert C_\alpha
\right\Vert h^\prime_{ij}, \chi^\prime\bigr\rangle = \int_{c_\alpha}
\delta g\, \delta\phi\, \Delta_\Phi [g]\, \delta \, [\Phi(g)] 
\exp\, (iS[g,\phi])\ \ . \tag sixone
$$
Here the sum is over four-dimensional metrics $g$ and field
configurations that induce the assigned three-metrics and fields on the
boundaries of $M$ and lie in the class $c_\alpha$. The functional $S$ is
the action for metric coupled to matter, and $\Phi$ is a suitable gauge
fixing condition.

\item{} Pure initial or final conditions are represented by wave
functions on the superspace of three-metrics and spatial matter field
configurations that satisfy operator versions of the four constraints of
general relativity
$$
{\cal H}_\mu ({\bf x})\, \Psi\, \left[h_{ij} ({\bf x}), \chi ({\bf
x})\right]=0\ .
\tag sixtwo
$$
One of these is the Wheeler-DeWitt equation. The action of the class
operators on such wave functions is specified by a linear (but not
necessarily positive) inner product $\circ$. Thus,
$$
\left\langle \Phi | C_\alpha | \Psi \right\rangle = \Phi^*\,
\bigr[h^{\prime\prime}_{ij}, \chi^{\prime\prime}\bigr]\, \circ\, \bigl\langle
h^{\prime\prime}_{ij}, \chi^{\prime\prime}\bigl\Vert C_\alpha \bigr\Vert
h^\prime_{ij}, \chi^\prime\bigr\rangle\,\circ \, \Psi\bigl[h^\prime_{ij},
\chi^\prime\bigr]\ . \tag sixthree
$$
There are various candidates for this $\circ$ product. The most immediate is
the analog of the Klein-Gordon inner product --- the DeWitt inner
product on hypersurfaces in superspace.

\item{} Density matrices representing non-pure initial and final
conditions may be constructed in linear spaces of wave functions
$\{\Psi_k[h_{ij},\chi]\}$ satisfying the constraints \(sixtwo). For example
$$
\bigl\langle h^{\prime\prime}_{ij}, \chi^{\prime\prime}\bigl\Vert \rho^i
\bigr\Vert h^\prime_{ij}, \chi^\prime\bigr\rangle= \sum\nolimits_k \Psi_k
\bigl[h^{\prime\prime}_{ij}, \chi^{\prime\prime}\bigr] p^\prime_k \Psi^*_k
\bigl[h^\prime_{ij}, \chi^\prime\bigr] \tag sixfour
$$
and similarly for $\rho^f$. The decoherence functional is then
constructed in the obvious way on these linear spaces
$$
D(\alpha^\prime, \alpha) = {\cal N}\, Tr\bigl[\rho^f C_{\alpha^\prime}
\rho^i C^\dagger_\alpha\bigr] \equiv {\cal N} \sum\nolimits_{ij}
p^{\prime\prime}_i \left\langle\Phi_i\left|C_{\alpha^\prime}\right|
\Psi_j\right\rangle\, p^\prime_i \left\langle\Psi_j\left|
C_\alpha\right| \Phi_i\right\rangle \tag sixfive
$$
where ${\cal N}$ is a normalizing factor ensuring
$\sum_{\alpha^\prime\alpha} D(\alpha^\prime, \alpha)=1$.

The generalized quantum mechanics defined by the above three elements 
does not require a fixed background spacetime or its associated causal
structure, at least formally. No special set of
spacelike surfaces has been singled out in the specification of sets of
fine-grained histories, the sets of possible alternatives, or the
amplitudes that define the decoherence functional. This theory of
four-dimensional quantum spacetimes is in fully four-dimensional form.

One cannot expect that this theory of quantum spacetime can be
reformulated in terms of states on spacelike surfaces, not least for the
reasons given in the Introduction. More specifically, this follows from
the nature of the fine-grained histories. Spacelike surfaces may be
defined intrinsically in a cosmological four-geometry by their 
 three-geometries including such properties as the total spatial volume.
However, there will be fine-grained quantum geometries in which surfaces
of a given three-geometry occur an arbitrary number of times. In this
sense, the histories of quantum general relativity move both ``forward''
and ``backward'' in any time defined by three-geometry. The
factorization of histories represented by relations like \(threefour)
cannot be expected to hold and states on spacelike surfaces in
superspace cannot be
defined.

Yet, the usual formalism of quantum mechanics with states on spacelike
surfaces in spacetime must hold in some approximation when geometry is 
approximately
fixed and classical. To see how this comes about consider a partition of
the fine-grained histories into coarse-grained classes $\{c_\alpha\}$
representing different possible classical and non-classical evolutions
defined to distance accuracies well above the Planck scale. Denote the possible
classical evolutions by $\{c_\gamma\}$ and suppose these are further refined
by partitions defining different behaviors of the matter fields in these
geometries into classes $\{c_{\gamma\beta}\}$. Suppose further that the
initial and final conditions are such that, when these classes are
sufficiently coarse grained, the integral over metrics from \(sixone) in
\(sixthree) may be done by the method of steepest descents with but a
single classical metric $g_\gamma$ contributing in each classical class
and no metrics at all in the non-classical classes. The class operator
matrix elements in this approximation vanish for the non-classical
classes and are determined by functional integrals of the form
$$
\int\nolimits_{c_{\gamma\beta}}\delta\phi\,
\exp\bigl(iS[g_\gamma,\phi]\bigr) \tag sixsix
$$
for those classes $c_{\gamma\beta}$ that define classical behavior of
geometry.  Functional integrals like \(sixsix) define a matter
field theory in a fixed background spacetime $g_\gamma$. These
amplitudes can be equivalently constructed from states of the matter
field on spacelike surfaces and their evolution. The classical metric
$g_\gamma$ fixes the meaning of ``spacelike'' and defines the notion(s)
of time.

In this way the usual formulation of quantum mechanics is recovered as
an approximation to the more general theory for those coarse-grainings
and initial and final conditions in which spacetime geometry is
approximately fixed, behaves
classically, and can define the necessary fixed causal structure.

{\topinsert
\centerline{\includegraphics[angle=90,width=8.0in]{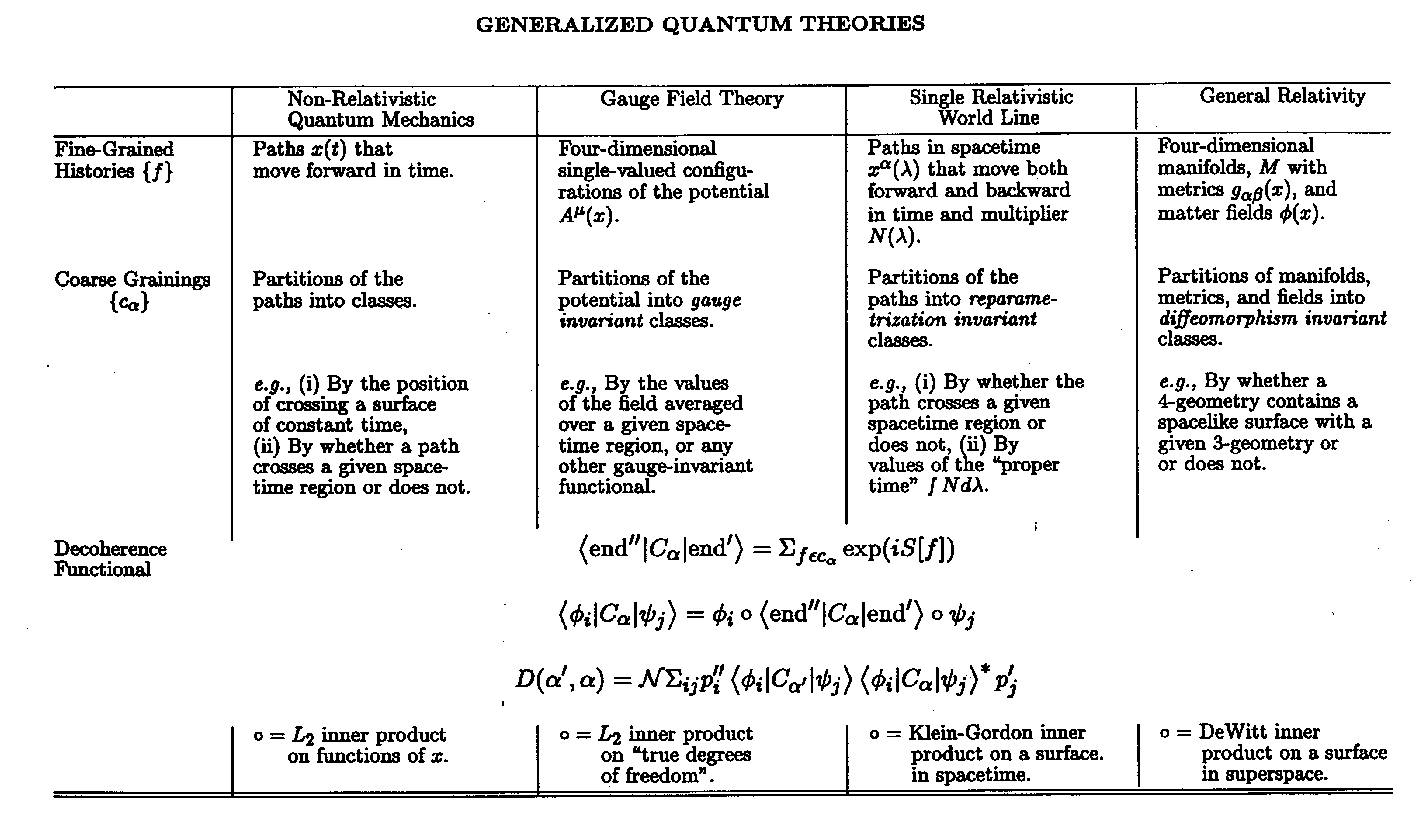}}
\endinsert}

\vskip .13 in
\leftline{\bf 7. Conclusions}

The accompanying table shows where we have been.  We have used the
framework of generalized quantum theory
 to  exhibit a series of generalizations of 
usual quantum theory. In each case we have specified the three elements: 
fine-grained histories, allowed coarse grainings, and decoherence
functional. The successive generalizations have incorporated spacetime
alternatives, gauge symmetries, and histories that move forward and
backward in  time --- all expected features of a theory of quantum
spacetime. We were able to build on these
generalizations to formally sketch a quantum theory of geometry and
matter fields that does not require a fixed background spacetime
geometry or causal structure  as familiar quantum theory does. 

>From this perspective the usual framework, with its states on spacelike
surfaces their unitary evolution and reduction, is not the most
fundamental formulation of quantum theory.  Rather it is an
approximation to a more general theory sketched above that is free from the
prerequisite of a fixed background spacetime and therefore applicable at
the Planck scale. 

\vskip .13 in
\leftline{\bf 8. Acknowledgment} 

The preparation of this report as well as the work described in it were
supported in part by the US National Science Foundation under grant
PHY90-08502.

\references

\gdef\journal#1, #2, #3, 1#4#5#6{{\sl #1} {\bf #2}
(1#4#5#6) #3}

\refis{Harpp} J.B.~Hartle, {\it Spacetime Quantum Mechanics and the
Quantum Mechanics of Spacetime} in {\sl
Gravitation and Quantizations}, Proceedings of the 1992 Les Houches
Summer School, edited by B. Julia and J. Zinn-Justin, Les Houches Summer
School Proceedings Vol. LVII, North Holland, Amsterdam (1995);
 UCSBTH-92-21, gr-qc/9304006.

\refis{Ish92} C.~Isham, in {\sl Recent Aspects of Quantum Fields},
ed.~by H.~Mitter and H.~Gausterer, Springer-Verlag, Berlin (1992).

\refis{Ish93} C.~Isham, in {\sl Integrable Systems, Quantum Groups,
and Quantum Field Theories}, ed by L.A.~Ibort and M.A.~
Rodriguez, Kluwer Academic Publishers, London (1993).

\refis{Kuc92} K.~Kucha\v r, in {\sl Proceedings of the 4th Canadian
Conference on General Relativity and Relativistic Astrophysics}, ed.~by
G.~Kunstatter, D.~Vincent, and J.~Williams, World Scientific, Singapore,
(1992).

\refis{Unr91} W.~Unruh, in {\sl Gravitation: A Banff Summer Institute},
ed.~by R.~Mann and P.~Wesson, World Scientific, Singapore (1991).

\refis{Harsum} J.B.~Hartle, in {\sl Directions in General Relativity,
Volume 1: A Symposium and Collection of Essays in honor of Professor
Charles W. Misner's 60th Birthday}, ed. by B.-L.~Hu,
M.P.~Ryan, and C.V.~Vishveshwara, Cambridge University Press, Cambridge
(1993); and in {\sl Gravitation and
Relativity
1989: Proceedings of the 12th International Conference on
General Relativity and Gravitation} ed.~by N.~Ashby, D.F.~Bartlett, and
W.~Wyss, Cambridge University Press, Cambridge (1990), 
gr-qc/\break
9210006 

\refis{Gri84} R.~Griffiths,  \journal J. Stat. Phys., 36, 219, 1984.

\refis{Omnsum} R.~Omn\`es, \journal J. Stat. Phys., 53, 893, 1988,
\journal ibid, 53, 933, 1988; \journal ibid, 53, 957, 1988; \journal
ibid, 57, 357, 1989; \journal Rev.~Mod.~Phys., 64, 339, 1992; {\sl
Interpretation of Quantum Mechanics}, (Princeton University Press,
Princeton, 1994).

\refis{GH90a} M.~Gell-Mann and J.B.~Hartle in {\sl Complexity, Entropy,
and the Physics of Information, SFI Studies in the Sciences of
Complexity}, Vol.  VIII, ed. by W. Zurek,  Addison Wesley, Reading, MA
or in {\sl Proceedings of the 3rd
International Symposium on the Foundations of Quantum Mechanics in the
Light of
New Technology} ed.~by S.~Kobayashi, H.~Ezawa, Y.~Murayama,  and
S.~Nomura,
Physical Society of Japan, Tokyo (1990).

\refis{Har91a}J.B.~Hartle, {\it The Quantum Mechanics of Cosmology}, in
{\sl
Quantum Cosmology and Baby Universes:  Proceedings of the 1989 Jerusalem
Winter
School for Theoretical Physics}, ed. by ~S.~Coleman, J.B.~Hartle,
T.~Piran,
and S.~Weinberg, World Scientific, Singapore (1991) pp. 65-157.

\refis{Har91b} J.B.~Hartle, \journal Phys. Rev., D44, 3173, 1991.

\refis{Ish94} C.J.~Isham, \journal J. Math. Phys., 35, 2157, 1994.

\endreferences

\endit
\end